\documentclass[preprint,showpacs,amsmath,amssymb]{revtex4}

\usepackage{graphicx,amsmath,amssymb,bm,multirow}
\usepackage{amsthm}
\usepackage{amsfonts}
\usepackage{dcolumn}
\usepackage{bm}
\usepackage[colorlinks,linkcolor=blue,anchorcolor=blue,citecolor=blue]{hyperref}
\usepackage[usenames,dvipsnames]{color}

\newcommand{\beq}{\vspace{0.5em}\begin{equation}}
\newcommand{\eeq}{\end{equation}\vspace{0.5em}}
\newcommand{\beqn}{\vspace{0.5em}\begin{eqnarray}}
\newcommand{\eeqn}{\end{eqnarray}\par\vspace{0.5em}\noindent}

\newcommand{\bsub}{\begin{subequations}}
\newcommand{\esub}{\end{subequations}}

\begin{document}
 \preprint{preprint}

\title{Beyond relativistic mean-field studies of low-lying states in neutron-deficient krypton isotopes}
\author{Y. Fu}
\affiliation{School of Physical Science and Technology, Southwest University, Chongqing 400715, China}
\author{H. Mei}
\affiliation{School of Physical Science and Technology, Southwest University, Chongqing 400715, China}
\author{J. Xiang}
\affiliation{School of Physical Science and Technology, Southwest University, Chongqing 400715, China}
\author{Z. P. Li}
\affiliation{School of Physical Science and Technology, Southwest University, Chongqing 400715, China}
\author{J. M. Yao}\email{jmyao@swu.edu.cn}
\affiliation{School of Physical Science and Technology, Southwest University, Chongqing 400715, China}
\affiliation{Physique Nucl\'eaire Th\'eorique, Universit\'e Libre de Bruxelles, C.P. 229, B-1050 Bruxelles, Belgium}
 \author{J. Meng}
\address{State Key Laboratory of Nuclear Physics and Technology,
  School of Physics, Peking University, Beijing 100871, China}
\address{School of Physics and Nuclear Energy Engineering, Beihang University, Beijing 100191, China}
\address{Department of Physics, University of Stellenbosch, Stellenbosch, South
Africa}
\begin{abstract}
\begin{description}
\item[Background] Neutron-deficient krypton isotopes are of particular interest due to the coexistence of oblate and prolate shapes in low-lying states and the transition of ground-state from one dominate shape to another as a function of neutron number. Moreover, the onset of large $E2$ transition strength around $^{76}$Kr indicates the erosion of $N=40$ sub-shell gap.

\item[Purpose] A detailed interpretation of these phenomena in neutron-deficient Kr isotopes  requires the use of a method going beyond a mean-field approach that permits to determine spectra and transition probabilities. The aim of this work is to provide a systematic calculation of low-lying state in the even-even $^{68-86}$Kr isotopes and to understand the shape coexistence phenomenon and the onset of large collectivity around $N=40$ from beyond relativistic mean-field studies.

\item[Method] The starting point of our method is a set of relativistic mean-field+BCS wave
    functions generated with a constraint on triaxial deformations $(\beta, \gamma)$. The excitation energies and electric multipole transition strengths of low-lying states are calculated by solving a five-dimensional collective Hamiltonian (5DCH) with parameters determined by the mean-field wave functions. To examine the role of triaxiality, a configuration mixing of both particle number (PN) and angular momentum (AM)  projected axially deformed states is also carried out within the exact generator coordinate method (GCM) based on the same energy density functional.

\item[Results] The energy surfaces, the excitation energies of $0^+_2, 2^+_1, 2^+_2$ states, as well as the $E0$ and $E2$ transition strengths are compared with the results of similar 5DCH calculations but with parameters determined by the non-relativistic mean-field wave functions, as well as with the available data. The results show a picture of oblate-triaxial-prolate shape transition. Coexistence of low-lying excited $0^+$ states is found to be a common feature in the neutron-deficient Kr isotopes. The underlying mechanism responsible for the shape coexistence is discussed.

\item[Conclusions] The main features of the low-spin spectra and the systematics of excitation energies and transition strengths in the neutron-deficient Kr isotopes are reproduced very well. The effects of dynamic correlations and triaxiality turn out to have important influences on the balance between the competing oblate and prolate states. An exact treatment of configuration mixing of PN+AM projected triaxial states is highly  demanded to pin down these effects.
\end{description}
\end{abstract}

\maketitle

 \section{\label{introduction}Introduction}

The low-lying states of neutron-deficient even-even krypton isotopes are of particular interest due to the rapid structural change with the neutron number and the presence of multiple shape coexistence, i.e. several $0^+$ states with different intrinsic shapes coexist at low excitation energy.
The irregularities observed in the ground-state bands at low spin in $^{74,76}$Kr was firstly suggested to be explained  by shape coexistence in Ref.~\cite{Piercey81}.
This interpretation is supported by the observation of a metastable low-lying $0^+_2$ state in $^{72,74}$Kr~\cite{Chandler97,Bouchez03} and the spectroscopic quadrupole moments for $^{74,76}$Kr~\cite{Clement07}. It shows clearly that the spectroscopic quadrupole moment of the ground-state band and that of the one based on the excited $0^+_2$ state have opposite signs, indicating the coexistence of a prolate ground-state with an oblate low-lying excited state.

Experimental evidence indicates a rapid structural change in neutron-deficient krypton isotopes. From $^{72}$Kr to $^{74,76}$Kr, a sudden increase of $B(E2: 0^+_1\to 2^+_1)$ value is interpreted as the transition from an oblate shape to a prolate one in the ground-state~\cite{Gade05}, which can also explain the rapid increase of charge radius~\cite{Angeli04}. From $^{76}$Kr up to $^{86}$Kr, the $B(E2: 0^+_1\to 2^+_1)$ value decreases smoothly, indicating the gradually decrease of collectivity towards neutron shell closure. The large collectivity around $^{76}$Kr implies the erosion of $N=40$ sub-shell gap.

Both shell models and mean-field based methods are adopted to interpret the complex and rapidly changing structure of nuclei in this mass region.

In the shell models, $np$-$mh$ excitations across the $N=40$ sub-shell are necessary to describe the neutron-deficient Kr isotopes, which is out of the reach of the conventional shell model. The complex excited VAMPIR approach using a modified $G$ matrix with effective charges~\cite{Petrovici00} or the Monte Carlo shell model using an effective quadrupole-plus-pairing residual interaction~\cite{Langanke03} has been carried out to analyze the low-lying excited states. The neutron-proton effective interaction and the excitation of nucleons across the $N=40$ sub-shell into $g_{9/2}$ orbital were claimed to be responsible for the shape mixing and the onset of large $B(E2)$ values respectively. Recently, the SD-pair shell model (SDPSM)~\cite{Chen97} with the $Z=28$, $N=50$ core was adopted to study the collective properties in the even-even $^{78-84}$Kr isotopes. The electromagnetic transitions among the low-lying states have been described rather well by a phenomenological Hamiltonian with parameters fitted to the excitation energies and by introducing effective charges for protons and neutrons~\cite{Wang10}.

In the framework of the self-consistent mean-field approaches, structural change and shape coexistence are described in terms of shapes and deformed shells~\cite{Meng05}. The presence of multiple local minima in the energy surface associated with sizable gaps in the single-particle spectrum are used to interpret the shape coexistence within the mean-field approximation. However, these isolated intrinsic shapes could be mixed with each other under the quantum fluctuation. A quantitative assessment of this mixing and a detailed description of the spectrum require ``going beyond the mean field" by means of symmetry restoration and taking into account fluctuations in the deformation degree of freedom.

Shape coexistence in the neutron-deficient $^{72-78}$Kr isotopes has been studied by the mixing of particle number and angular momentum projected (GCM+PNP+1DAMP) axially deformed Hartree-Fock-Bogolibov (HFB) states with Skyrme force SLy6 in Ref.~\cite{Bender06}. An oblate ground-state band coexisting with a prolate excited band was obtained for all light Kr isotopes. The failure in predicting the energy order of the oblate and prolate states could be due to the following reasons:
1) One is the deficiency of the SLy6 {force} for the {\em fp} shell nuclei.
A prolate ground-state for $^{74,76}$Kr requires a reduction of the gap at $N = 40$ and the separation between the $f_{5/2}$ and the $p_{3/2}$ levels~\cite{Bender06}.
Besides, as pointed out in Ref.~\cite{Reinhard99}, the balance between the prolate and the oblate state can also be altered with a different surface energy coefficient in the Skyrme force;
2) The other is the limitation to axial states in the calculations. Recently, triaxiality was shown to be crucial to reproduce the transition from the oblate to the prolate shape from $^{72}$Kr to $^{74,76}$Kr based on the comparison between the collective Hamiltonian calculations with and without triaxial states for these nuclei using the D1S force~\cite{Girod09}.

In the past decades, the relativistic mean-field (RMF) theory, which relies on the basic ideas of effective field theory and density functional theory and is therefore referred to as single-reference covariant density functional theory (SR-CDFT), has achieved great success in the description of ground-state properties of both spherical and deformed nuclei all over the nuclear chart~\cite{Reinhard89,Ring96,Vretenar05,Meng06}. In particular, the energy density functional of point-coupling type has recently attracted more and more attention~\cite{Niksic11}. It shows a great advantage in the extension for nuclear low-lying excited states by implementing projection techniques~\cite{Yao08,Yao09} and generator coordinate method (GCM)~\cite{Niksic06,Yao10,Yao11,Yao11-C}. This types of implementation is also referred to as multi-reference (MR) CDFT.  Since {an} exact MR-CDFT calculation for triaxially deformed nuclei requires the mixing of symmetry restored triaxial states, the numerical computation is presently very expensive for a systematical study of nuclear low-lying states in medium and heavy-mass regions. As a Gaussian overlap approximation of the exact MR-CDFT, the model of five-dimensional collective Hamiltonian (5DCH) with parameters determined from the mean-field calculation is much less numerical demanding and turns out to be a powerful tool for the systematical studies of nuclear low-lying states~\cite{Libert99,Prochniak04,Niksic09,Li09,Li10,Li11,Li12,Niksic11}. Within this framework, the rapid shape evolution in the neutron-rich Kr isotopes around  $N=60$~\cite{Xiang12} have been studied {using the PC-PK1 force~\cite{Zhao10} and the results were compared with the similar 5DCH calculations but using the Skyrme force SLy4~\cite{Mei12}}.

In this paper we extend our studies to the neutron-deficient Kr isotopes around $N=40$. A systematic beyond-mean-field calculation of the low-lying states will be carried out to examine the validity of the CDFT for nuclei with shape coexistence structure, and explore the mechanism for the shape coexistence, the onset of large collectivity around $N=40$, and the role of triaxiality.

The paper is arranged as follows. In Sec.~\ref{Sec.II}, we present a brief introduction to the 5DCH that is used to calculate the low-lying states in neutron-deficient Kr isotopes. In Sec.~\ref{Sec.III}, the energy surfaces, low-energy excitation spectra, electric monopole and quadrupole transition strengths are presented and discussed in comparison with the results of the 5DCH calculation using the non-relativistic Gogny force D1S and available data. Taking $^{76}$Kr as an example, the spectra will be also compared with the exact GCM+PNP+1DAMP calculations using the PC-PK1 force and the SLy6 force. A summary of our findings and conclusions is given in Sec.~\ref{Sec.IV}.

 \section{Sketch of the theoretical framework}%
 \label{Sec.II}
The quantized 5DCH that describes the nuclear excitations of quadrupole vibration, rotation, and their couplings can be written in the form~\cite{Libert99,Prochniak04,Li09}
\begin{equation}
\label{hamiltonian-quant}
\hat{H} =
\hat{T}_{\textnormal{vib}}+\hat{T}_{\textnormal{rot}}
              +V_{\textnormal{coll}} \; ,
\end{equation}
where $V_{\textnormal{coll}}$ is the collective potential,
\begin{equation}
\label{Vcoll}
 {V}_{\textnormal{coll}}(\beta,\gamma)
 = E_{\textnormal{tot}}(\beta,\gamma)
  - \Delta V_{\textnormal{vib}}(\beta,\gamma) - \Delta
  V_{\textnormal{rot}}(\beta,\gamma),
\end{equation}
with $E_{\textnormal{tot}}(\beta,\gamma)$ the standard nuclear total energy in the mean-field calculation.  The $\Delta V_{\textnormal{vib}}$ and $\Delta V_{\textnormal{rot}}$ are zero-point-energy (ZPE) of vibrational and rotational motions, respectively, calculated with the cranking approximation~\cite{Li09}. The vibrational kinetic energy reads,
\begin{eqnarray}
\hat{T}_{\textnormal{vib}}
 &=&-\frac{\hbar^2}{2\sqrt{wr}}
   \left\{\frac{1}{\beta^4}
   \left[\frac{\partial}{\partial\beta}\sqrt{\frac{r}{w}}\beta^4
   B_{\gamma\gamma} \frac{\partial}{\partial\beta}\right.\right.\nonumber\\
  && \left.\left.- \frac{\partial}{\partial\beta}\sqrt{\frac{r}{w}}\beta^3
   B_{\beta\gamma}\frac{\partial}{\partial\gamma}
   \right]+\frac{1}{\beta\sin{3\gamma}} \left[
   -\frac{\partial}{\partial\gamma} \right.\right.\nonumber\\
  && \left.\left.\sqrt{\frac{r}{w}}\sin{3\gamma}
      B_{\beta \gamma}\frac{\partial}{\partial\beta}
    +\frac{1}{\beta}\frac{\partial}{\partial\gamma} \sqrt{\frac{r}{w}}\sin{3\gamma}
      B_{\beta \beta}\frac{\partial}{\partial\gamma}
   \right]\right\},\nonumber\\
 \end{eqnarray}
and rotational kinetic energy,
\begin{equation}
\hat{T}_{\textnormal{\textnormal{\textnormal{rot}}}} =
\frac{1}{2}\sum_{k=1}^3{\frac{\hat{J}^2_k}{\mathcal{I}_k}},
\end{equation}
with $\hat{J}_k$ denoting the $k$-component of the angular momentum in
the body-fixed frame of a nucleus. It is noted that the mass
parameters $B_{\beta\beta}$, $B_{\beta\gamma}$, $B_{\gamma\gamma}$,
as well as the moments of inertia $\mathcal{I}_k$, depend on the
quadrupole deformation variables $\beta$ and $\gamma$,
\begin{equation}
\mathcal{I}_k = 4B_k\beta^2\sin^2(\gamma-2k\pi/3),~~ k=1, 2, 3.
\end{equation}
Two additional quantities that appear in the expression for the
vibrational energy: $r=B_1B_2B_3$, and
$w=B_{\beta\beta}B_{\gamma\gamma}-B_{\beta\gamma}^2 $, determine the
volume element in the collective space. The corresponding eigenvalue
problem is solved by means of expansion of eigenfunctions in terms of a
complete set of basis functions that depend on the deformation
variables $\beta$ and $\gamma$, and the Euler angles $\phi$,
$\theta$ and $\psi$~\cite{Pro.99}.

The dynamics of the 5DCH is governed by the seven functions of the intrinsic deformations $\beta$ and $\gamma$: the collective potential $V_{\rm coll}$, the three mass parameters:
$B_{\beta\beta}$, $B_{\beta\gamma}$, $B_{\gamma\gamma}$, and the
three moments of inertia $\mathcal{I}_k$. These functions are determined
by a set of mean-field wavefunctions, generated by the constrainted SR-CDFT calculations
on the mass quadrupole moments $q_{20}$ and $q_{22}$, which are related to $\beta$ and $\gamma$ by
 \bsub%
 \beqn%
 q_{20}&=& \sqrt{\frac{5}{16\pi}}\langle 2z^2-x^2-y^2\rangle = \frac{3}{4\pi}AR^2_0\beta\cos
 \gamma,~~~~~~~~~~~\\
 q_{22}&=& \sqrt{\frac{15}{32\pi}}\langle x^2-y^2\rangle = \frac{3}{4\pi}AR^2_0 \frac{1}{\sqrt{2}}\beta\sin\gamma,
 \eeqn
 \esub%
 where $R_0=1.2A^{1/3}$ fm, $A$ being the mass number.

 In the constrained SR-CDFT calculation, parity ($\hat P$), $x$-simplex symmetry ($\hat Pe^{i\pi\hat J_x}$), and time-reversal invariance are imposed for the single-particle states. The Dirac equation is solved by expanding in the basis of eigenfunctions of a three-dimensional harmonic oscillator in Cartesian coordinate with 12 major shells, which are found to be sufficient to obtain reasonably convergent results for the nuclei in this mass region. If not mentioned explicitly, the pairing correlations are treated by using the BCS approximation with a pairing force separable in momentum space~\cite{Tian09}. The numerical details of the calculations can also be found in Refs.~\cite{Niksic09,Li09,Li10,Li11,Li12,Xiang12,Mei12}.

 \section{Results and discussions}
 \label{Sec.III}
\begin{figure}[]
\centering
\includegraphics[clip=,width=8.5cm]{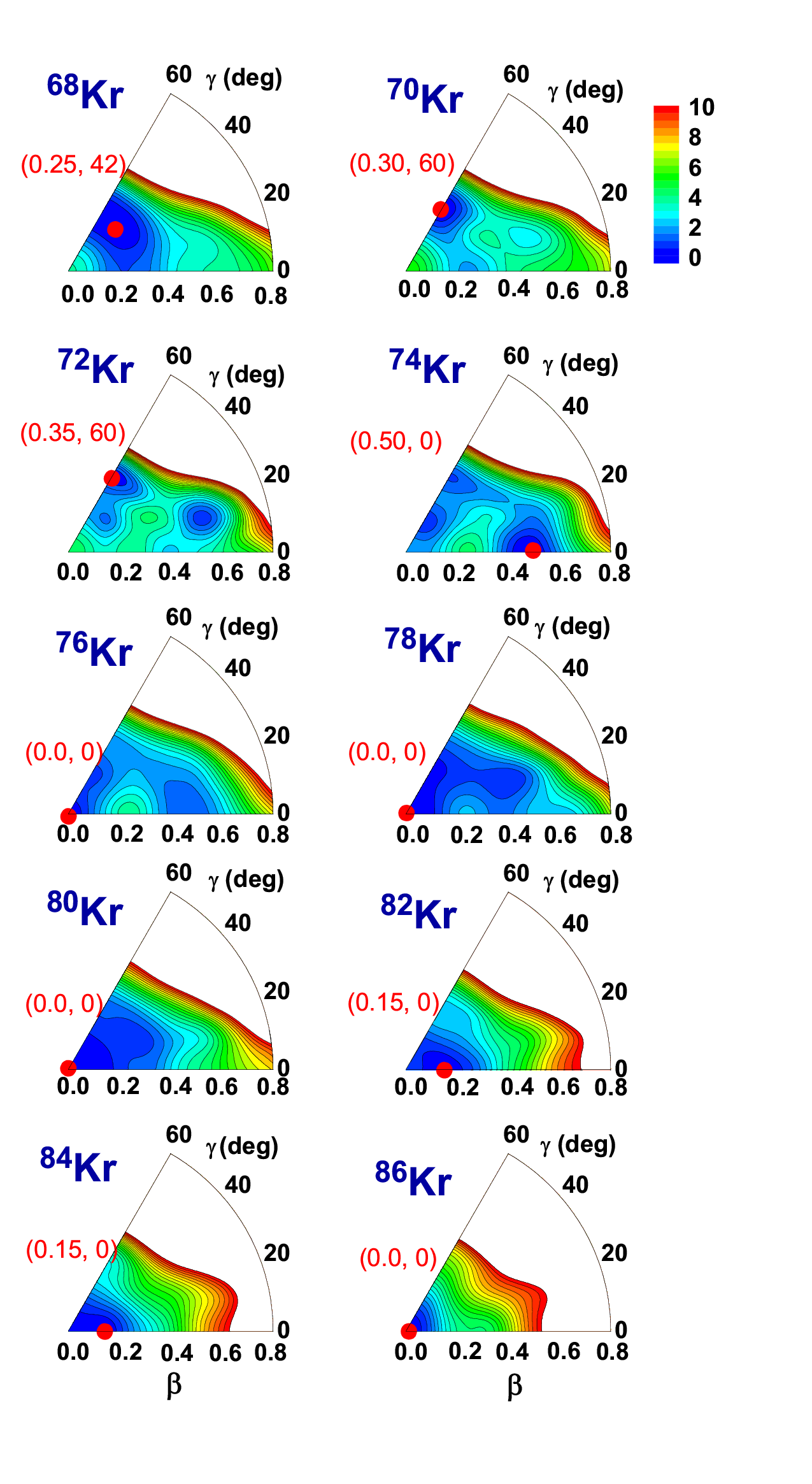}\vspace{-1cm}
\caption{(Color online) Energy surfaces of even-even $^{68-86}$Kr isotopes in the $\beta$-$\gamma$ plane from a constrained relativistic mean-field (RMF) plus BCS calculation using the PC-PK1 force. All energies are normalized to the absolute minimum, indicated with red bullets and corresponding $(\beta, \gamma)$ values. The energy difference between neighboring contour lines is 0.5~MeV.}
\label{fig:PESKr}
\end{figure}

\subsection{Energy surfaces}

 Figure~\ref{fig:PESKr} displays the energy surface in the $\beta$-$\gamma$ plane for the even-even $^{68-86}$Kr. Similar results have been obtained in the calculations with the D1S parametrization of Gogny force~\cite{Delaroche10}. There is a triaxial minimum  with $(\beta, \gamma)=(0.25, 42^\circ)$ in $^{68}$Kr. However, this minimum is soft along $\gamma$ direction and moves to the oblate side in $^{70}$Kr. Besides, a triaxial shoulder around $(0.55, 18^\circ)$ is developing in $^{70}$Kr and it becomes a local minimum in $^{72}$Kr, in which, the oblate minimum becomes fragmental and is separated into two minima with slightly smaller and larger deformations. An abrupt structural change is shown when the neutron number increases from $N=36$ to $38$. The triaxial shoulder moves to the prolate side around $(0.50, 0^\circ)$, which becomes the global minimum in $^{74}$Kr. This prolate minimum persists in $^{76}$Kr, but later shifts to the triaxial state around ($0.40, 18^\circ$) in $^{78}$Kr and disappears in the nuclei beyond $N=42$.

 It is noted that the global minimum in the energy surface is shifted to the spherical shape at $N=40$ with about 1.1 MeV lower in energy than the prolate minimum. However, the dynamic correlation at the beyond-mean-field level could reduce significantly the energy difference and leads to a largely deformed ground-state in $^{76}$Kr. This effect will be discussed in detail subsequently.  As the neutron number approaches $N=50$, the shape of ground-state becomes spherical. In short, the topography of energy surfaces displays an oblate-triaxial-prolate-spherical shape transition picture in the neutron-deficient Kr isotopes. The energy surfaces by the HFB calculations using the Gogny D1S force present a similar picture~\cite{Delaroche10}.

 \subsection{Systematics of low-lying states}

\begin{figure}[]
\centering
\includegraphics[width=7.5cm]{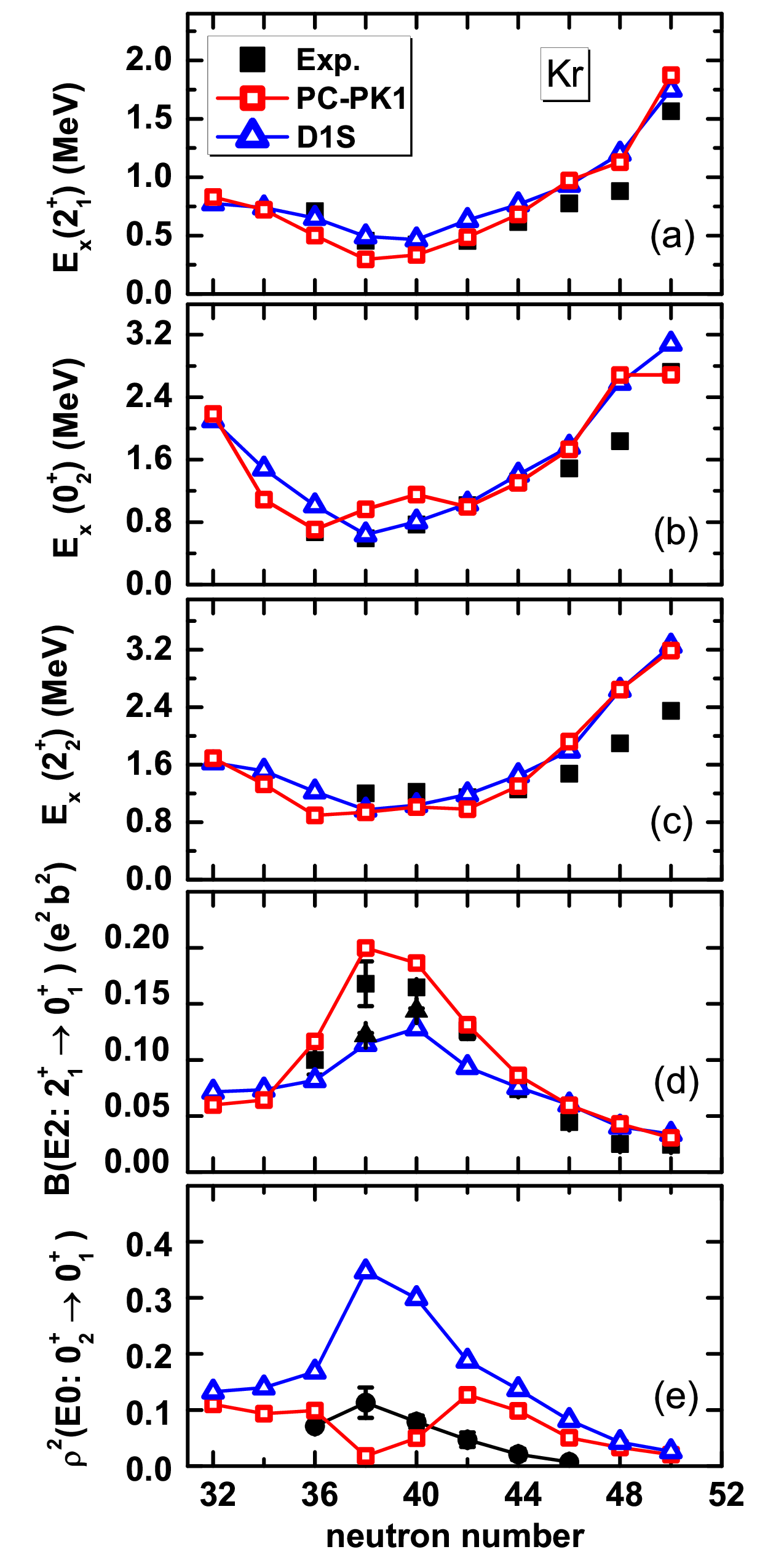}
\caption{(Color online) Excitation energies for $2^+_1$, $2^+_2$, and $0^+_2$ states, as well as the $B(E2: 2^+_1\to 0^+_1)$ and  $\rho^2(E0: 0^+_2\to 0^+_1)$  values as a function of neutron number in neutron-deficient Kr isotopes, in comparison with the similar calculations with the Gogny D1S force~\cite{Delaroche10} and available data taken from Ref.~\cite{NNDC} (squares), Ref.~\cite{Clement07} (triangles) and Ref.~\cite{Kibedi05} (circles).
}
\label{fig:KrBE2}
\end{figure}

  Figure~\ref{fig:KrBE2} displays the calculated 5DCH excitation energies for $2^+_1$, $0^+_2$, and $2^+_2$  states, as well as the $B(E2: 2^+_1\to 0^+_1)$ values as functions of the neutron number in the neutron-deficient Kr isotopes using the PC-PK1 force, in comparison with available data~\cite{NNDC}. In Ref.~\cite{Girod09}, similar calculations have been done using the Gogny D1S force. The Gogny D1S force has been adopted for a global study of the low-lying states for nuclei with proton numbers ranging from $Z=10$ to $Z=110$ and neutron numbers $N\leq200$~\cite{Delaroche10}. Therefore, the corresponding results are plotted as well for comparison.

 The 5DCH calculations based on the PC-PK1 force give similar results to those based on the D1S force, both of which reproduce the systematics rather well, in particular for the onset of large $B(E2:2^+_1\to 0^+_1)$ value around $N=40$. Quantitatively, there are some differences in the prediction for the nuclei around $N=40$. The PC-PK1 force presents a more dramatic evolution trend from $N=34$ to $N=38, 40$. Compared with the results of the PC-PK1 force, a lower $0^+_2$ state and a smaller $B(E2:2^+_1\to 0^+_1)$ value were obtained in the calculations using the D1S force. These differences are probably due to the strong configuration mixing in the calculation of the D1S force.

 An indicator of configuration mixing is the $E0$ transition strength,
 \begin{equation}
  \rho^2(E0: 0^+_2 \to 0^+_1)
  =\left|\frac{\langle0^+_2|\sum_k e_kr_k^2|0^+_1\rangle}   {eR^2_0}\right|^2,
 \end{equation}
 where $R_0=1.2A^{1/3}$~fm. In general, a large $E0$ transition strength is a signature of strong mixing of states associated with the coexisting shapes~\cite{Heyde11}. The panel (e) of Fig.~\ref{fig:KrBE2}  shows the value of $\rho^2(E0: 0^+_2\to 0^+_1)$ as a function of neutron number. A reasonable agreement between the two 5DCH calculated results is observed in all the neutron-deficient Kr isotopes, except for the nuclei around $N=40$ again. The calculation using the D1S force overestimates the $\rho^2(E0: 0^+_2 \to 0^+_1)$ value in $^{74,76}$Kr by a factor of about 3, while the  PC-PK1 calculated results underestimates this value in $^{74}$Kr.

\begin{figure}[]
\centering
\includegraphics[width=8.5cm]{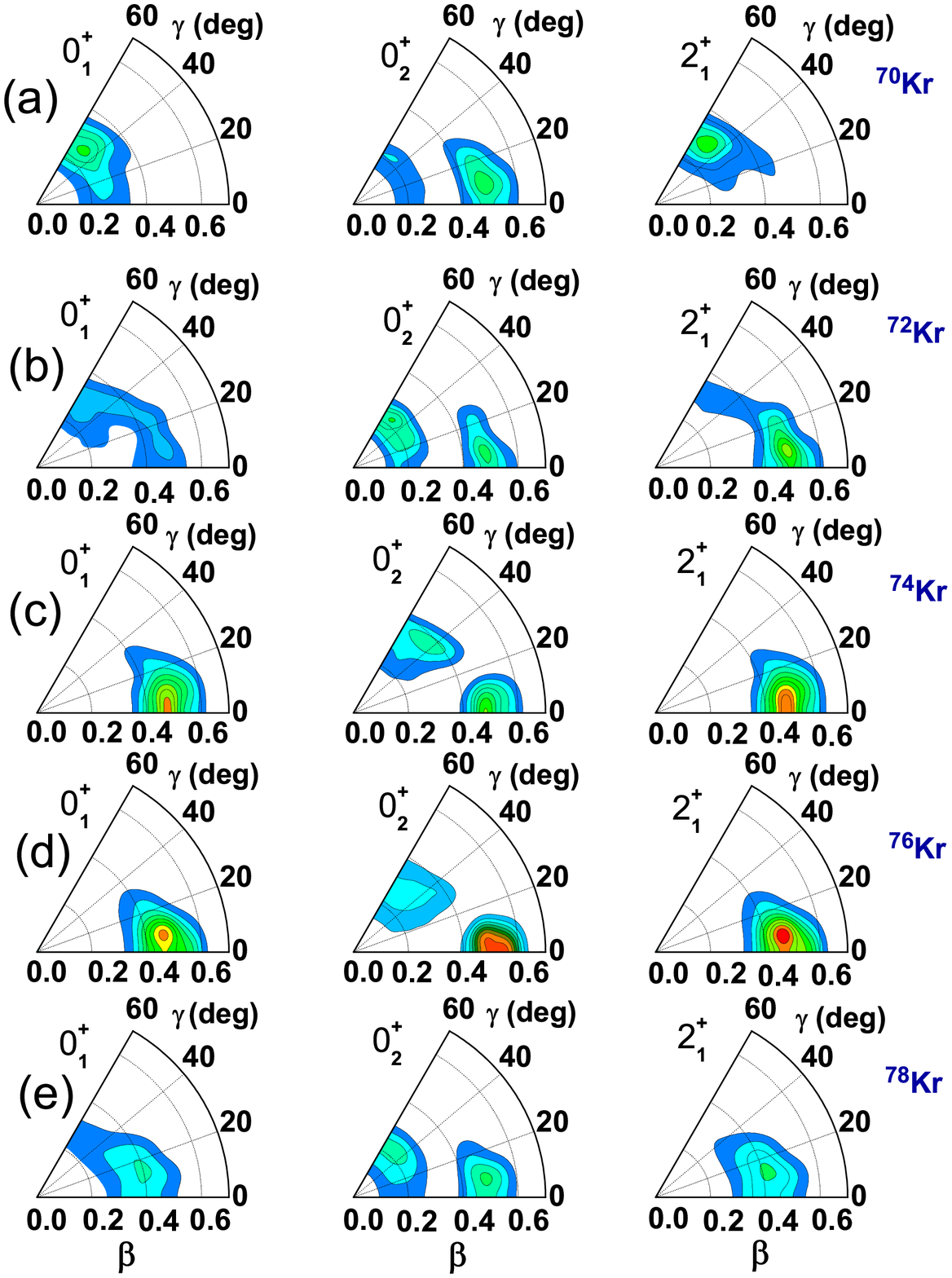}\vspace{-0.5cm}
\caption{(Color online) Distribution of the probability density $\rho_{J\alpha}(\beta, \gamma)$ for the first two $0^+$ states  and $2^+_1$ state in $^{70-78}$Kr.}
\label{wfs}
\end{figure}

 The detailed information about the structure of low-lying states is provided by the collective wave functions. Figure~\ref{wfs} displays the distribution of probability density $\rho_{J\alpha}$ in the $\beta$-$\gamma$ plane for the first two $0^+$ states and the $2^+_1$ state in $^{70-78}$Kr, where the $\rho_{J\alpha}$ satisfies the following normalization condition,
  \beq
  \int^\infty_0 \beta d\beta  \int^{2\pi}_0 \sin3\gamma d\gamma \rho_{J\alpha}(\beta,\gamma)=1.
  \eeq
  As the neutron number increases, the predominate configuration of ground-state is changing from a weakly deformed shape ($\vert\beta\vert=0.25$) at $N=34$ to a large prolate deformed one ($\vert\beta\vert=0.50$) at $N=38, 40$. The evolution of the predominate configuration shows again the onset of the large collectivity in the nuclei around $N=40$. The second $0^+$ state is predominately prolate deformed at $N=34$ and becomes a strong mixing of oblate and triaxial shapes at $N=36$. The distribution of probability density $\rho_{J\alpha}(\beta,\gamma)$ indicates that the $0^+_2$ state in $^{70,72}$Kr belongs to a $\beta$-vibration state. However, the $0^+_2$ state in $^{74,76}$Kr is a $\gamma$-vibration state and becomes a $\beta$-vibration state again in $^{78}$Kr. It can be understood from the energy surfaces shown in Fig.~\ref{fig:PESKr} that the $^{74,76}$Kr are soft under the distortion along $\gamma$-direction. In most cases, the distribution of probability density  of the $2^+_1$ state is similar to that of $0^+_1$ state, except that the $2^+_1$ state has already become predominantly prolate deformed at $N=36$. This picture is slightly different from that by the D1S force~\cite{Girod09}, which gives a predominantly oblate shape at $N=36$.

\subsection{Shape coexistence and Nilsson diagrams}

\begin{figure}[]
\centering
\includegraphics[width=9cm]{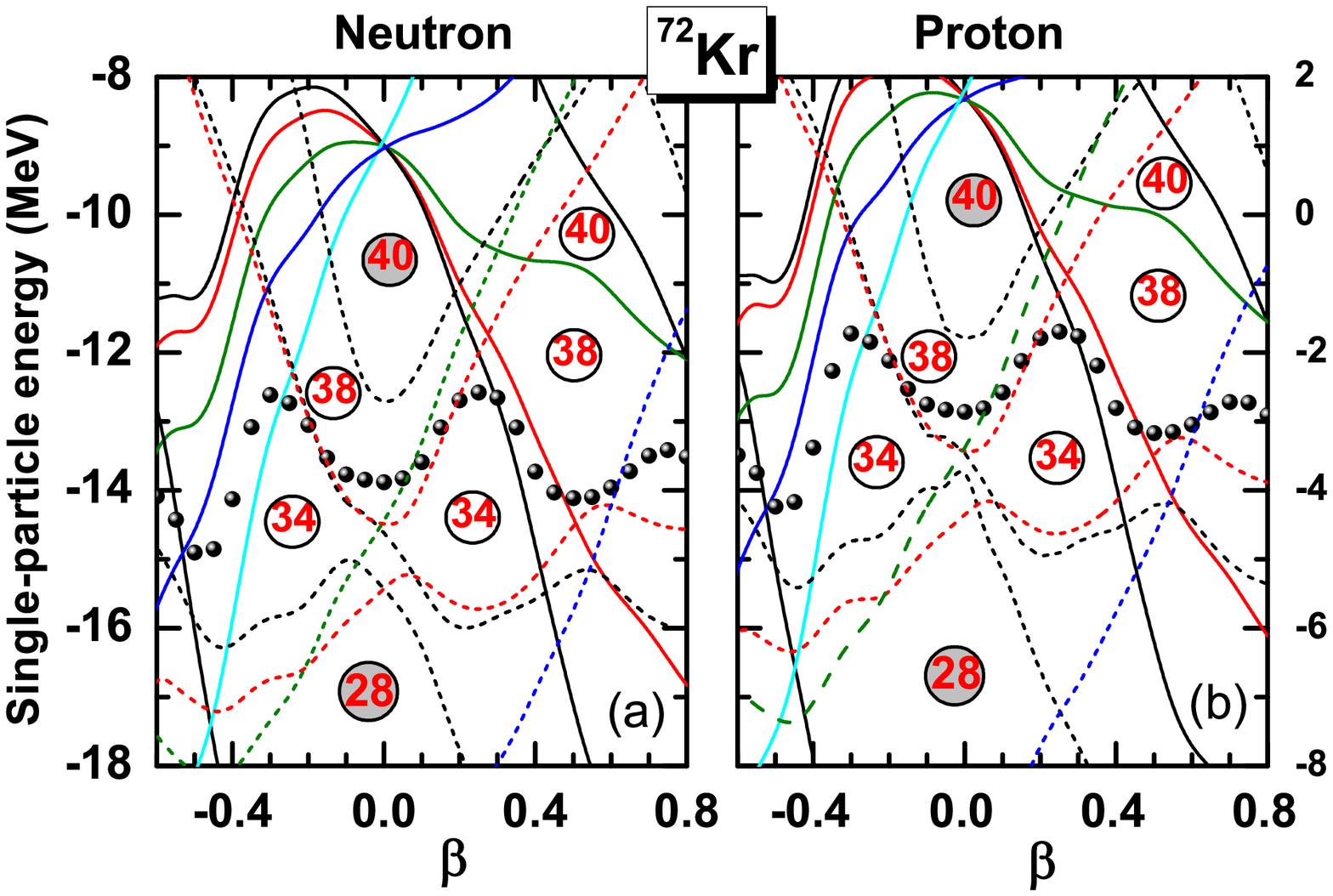}
\caption{(Color online) Nilsson diagram of the eigenvalues of the single-particle Hamiltonian for
neutrons and protons in $^{72}$Kr  as a function of the axial deformation parameter $\beta$, obtained with the PC-PK1 force. Fermi energies are indicated with bullets.}
\label{spe}
\end{figure}

Shape mixing is an important quantum concept in the study of shape coexistence in atomic nuclei. In order to have near-degenerate collective states with the same quantum numbers but different deformations, there has to be a mechanism to prevent their mixing. The sizable shell gaps in Nilsson diagram and large barriers between the minima in the energy surface is not sufficient to prevent their mixing. As discussed in Ref.~\cite{Yao12}, the intruder states with different parity can provide such mechanism. Therefore, shape coexistence is found mostly in nuclei with neutron or proton number around shell-closure and sub-shell closure and with the presence of intruder states~\cite{Heyde11}.

To explore the mechanism responsible for the shape coexistence in the neutron-deficient Kr isotopes, we plot in Fig.~\ref{spe} the single-particle energies of neutrons and protons as functions of quadrupole deformation parameter $\beta$. As already shown in Ref.~\cite{Bender06}, there are several sizable gaps located at different deformation regions in the Nilsson diagram of neutrons from $N=34$ to $N=40$.

We note that with increasing of deformation $\beta$, the downsloping $K = 1/2, 3/2$ ( or $K=9/2,7/2$) levels from the $1g_{9/2}$ orbital are diving into the Fermi sea in the prolate (or oblate) side and becoming intruder levels. Two large $N=34$ gaps are located in both prolate and oblate sides with a similar $\vert\beta\vert$ value around 0.25, which gives rise to the mixing of a weakly oblate configuration with a weakly prolate one in the ground-state of $^{70}$Kr. Since the intruder $\nu1g_{9/2}$ orbital becomes occupied in the region with $\vert\beta\vert>0.4$, the mixing between configurations with $\vert\beta\vert$ below and above 0.4 is weak, but not zero because of pairing effect. Therefore, the wave function of ground-state is concentrated in the small deformed region $(\vert\beta\vert= 0.25)$, while the second $0^+$ state is around the large deformed region $(\vert\beta\vert= 0.50)$. As a consequence, the mixing between the first two $0^+$ states in $^{70}$Kr is weak, as shown in Fig.~\ref{wfs}. In $^{72}$Kr, the intruder $\nu1g_{9/2}$ orbital is already occupied when the deformation reaches to $\vert\beta\vert\approx0.2$, which is smaller than the $\vert\beta\vert$ value of the state in the global energy minimum. As a result, there is a strong mixing between the two configurations corresponding to the global minimum and the second triaxial minimum respectively. There are two $N=38$ shell gaps. One is located at the spherical shape, extending to the oblate side up to $\vert\beta\vert\approx0.3$, the other is located in the prolate side at $\beta\approx0.5$. Since the prolate gap is much larger than the spherical-oblate one and the number of occupied intruder states is different in these two regions, the ground-state of $^{74}$Kr is finally concentrated around $\beta=0.5$ (cf. Fig.~\ref{wfs}). The mixing between the first two $0^+$ states is weak, which is however not supported by the data of $\rho^2(E0:0^+_2\to 0^+_1)$.

 \subsection{Spectroscopy of low-lying states in $^{70-78}$Kr}

\begin{figure}[]
\centering
\includegraphics[width=8cm]{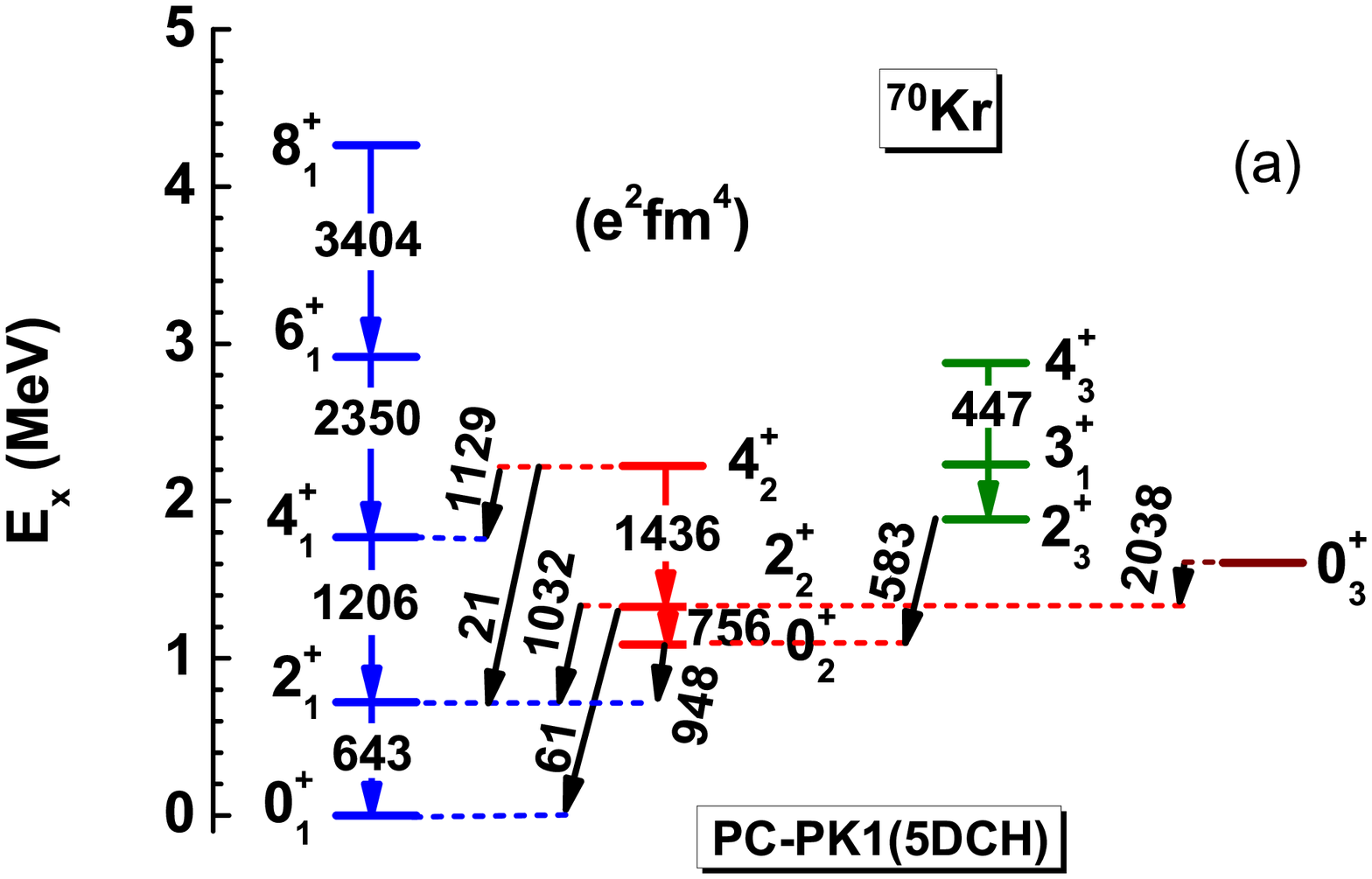}
\includegraphics[width=8cm]{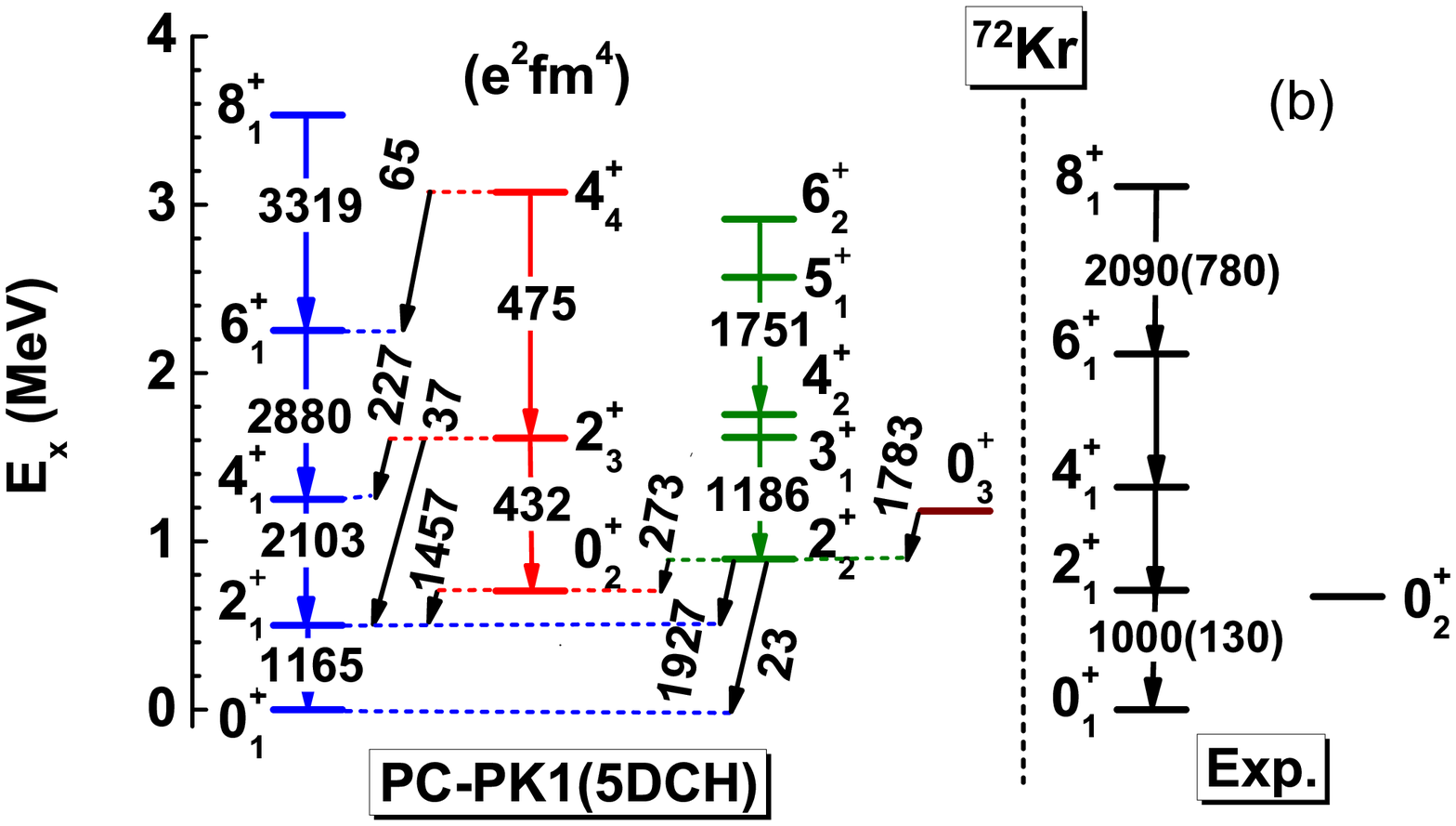}
\caption{(Color online) Low-spin spectra of $^{70,72}$Kr in comparison with available data. The $B(E2)$ transition strengths are given in $e^2$ fm$^4$. Experimental data are taken from Refs.~\cite{Angelis97,Gade05}.}
\label{fig:70-72Kr}
\end{figure}

\begin{figure*}[]
\centering
\includegraphics[width=12cm]{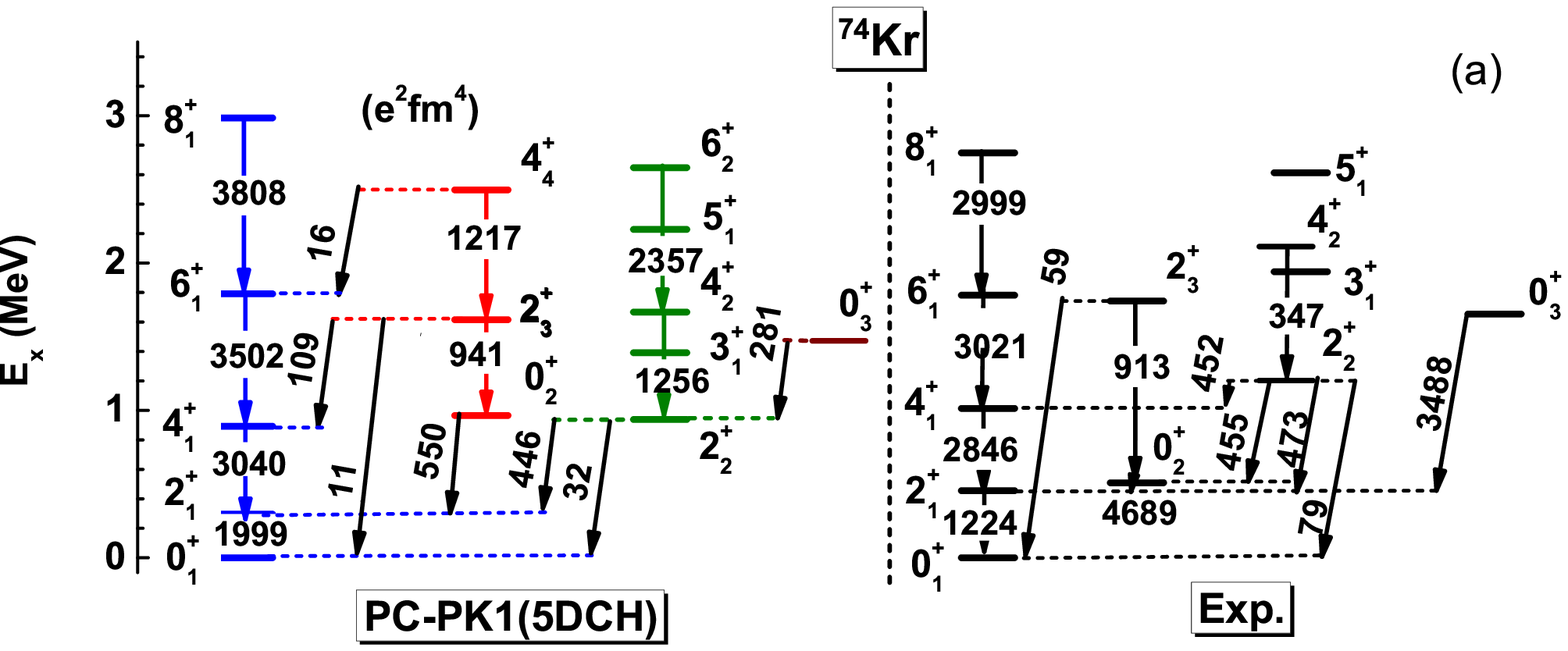}
\includegraphics[width=12cm]{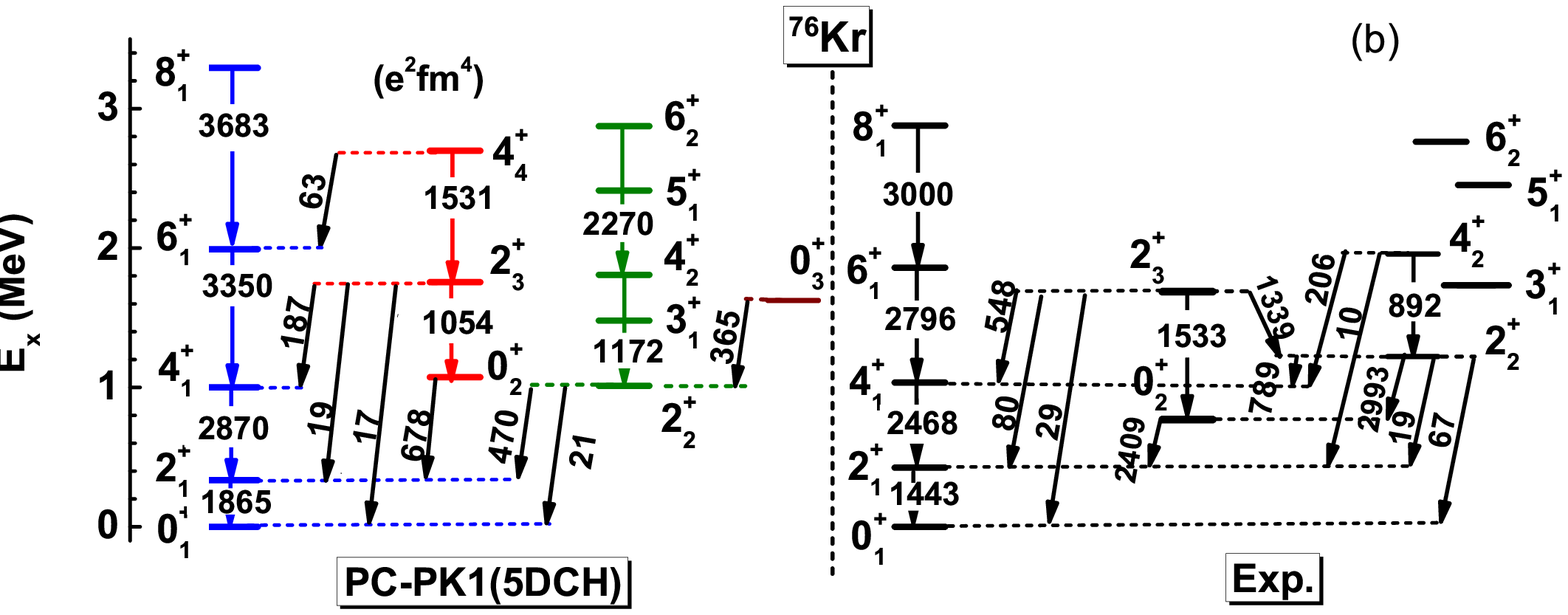}
\includegraphics[width=12cm]{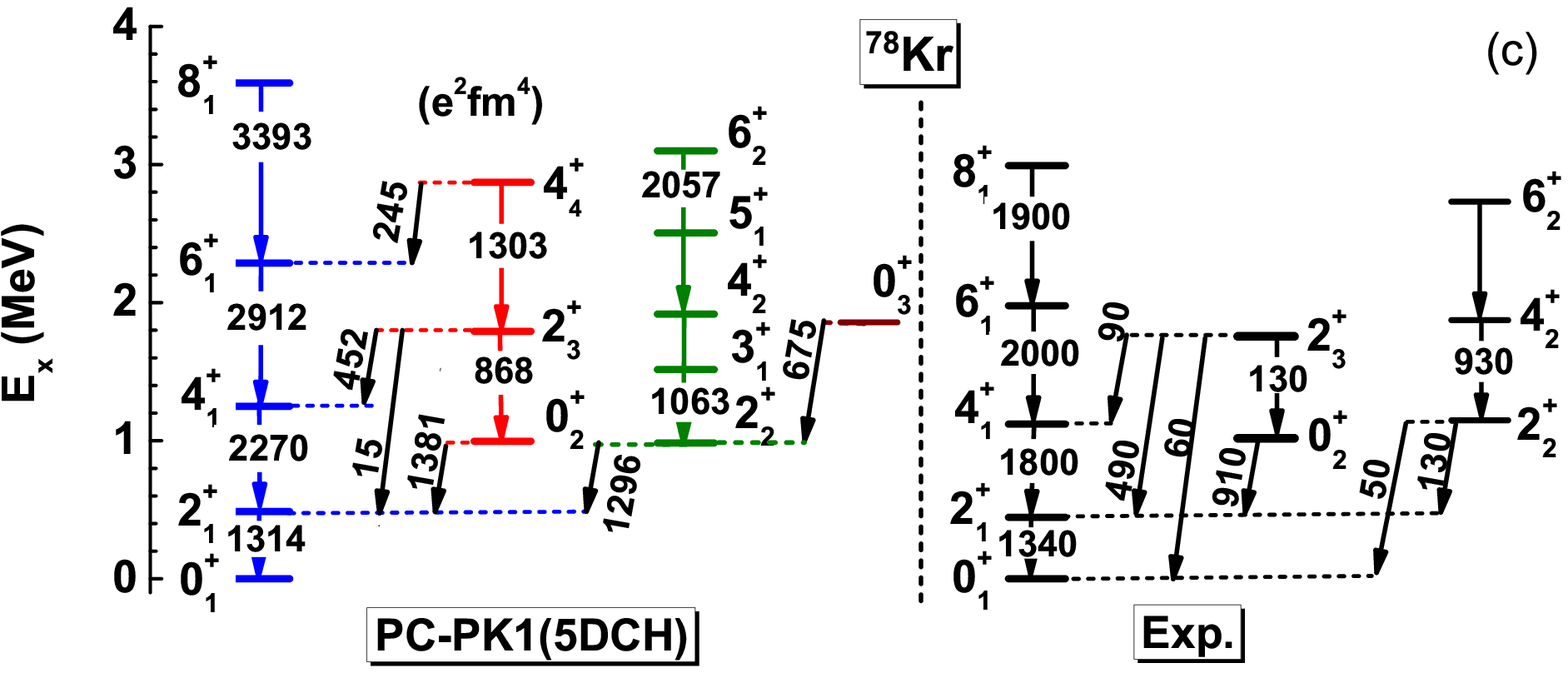}
\caption{(Color online) Same as Fig.~\ref{fig:70-72Kr}, but for $^{74,76,78}$Kr in comparison with available data taken from Refs.~\cite{Clement07,NNDC}.}
\label{fig:74-78Kr}
\end{figure*}
The 5DCH calculated low-spin spectra of $^{70-78}$Kr are compared with available data in Fig.~\ref{fig:70-72Kr} and Fig.~\ref{fig:74-78Kr}. The main features of the low-spin spectra are reproduced very well, in particular for the ground-state band and the low-lying $0^+_{2}$ state. The observed large $E2$ transition strength from the $0^+_2$ state to the $2^+_1$ state in $^{74,76}$Kr was reproduced in the 5DCH calculations using the D1S force~\cite{Clement07}, but is underestimated in our calculations by one order of magnitude. It is shown in Figs.~\ref{fig:70-72Kr} and~\ref{fig:74-78Kr} that the calculated $B(E2: 0^+_2\to 2^+_1)$ value in  $^{74,76}$Kr by the PC-PK1 force plus the separable pairing force is only half of the value of the neighboring nuclei, which corresponds to the weaker $E0$ transition strength, cf. Fig.\ref{fig:KrBE2}. The 5DCH calculations using the SLy6 force (or using the PC-PK1 force) together with the zero-range pairing force give much stronger $E0$ transition with $\rho^2(E0:0^+_2\to 0^+_1)=0.148$ (or $0.153)$ and larger $B(E2: 0^+_2\to 2^+_1)$ value for $^{76}$Kr. All of these results imply that the reproduction of large $E2$ transition strengths between the $2^+_1$ and $0^+_2$ states requires a strong mixing between the first two $0^+$ states, as it is obtained in the 5DCH calculations based on the D1S force.

 \subsection{Effects of dynamic correlations and triaxiality in $^{76}$Kr}

\begin{figure}[]
\centering
\includegraphics[width=8.5cm]{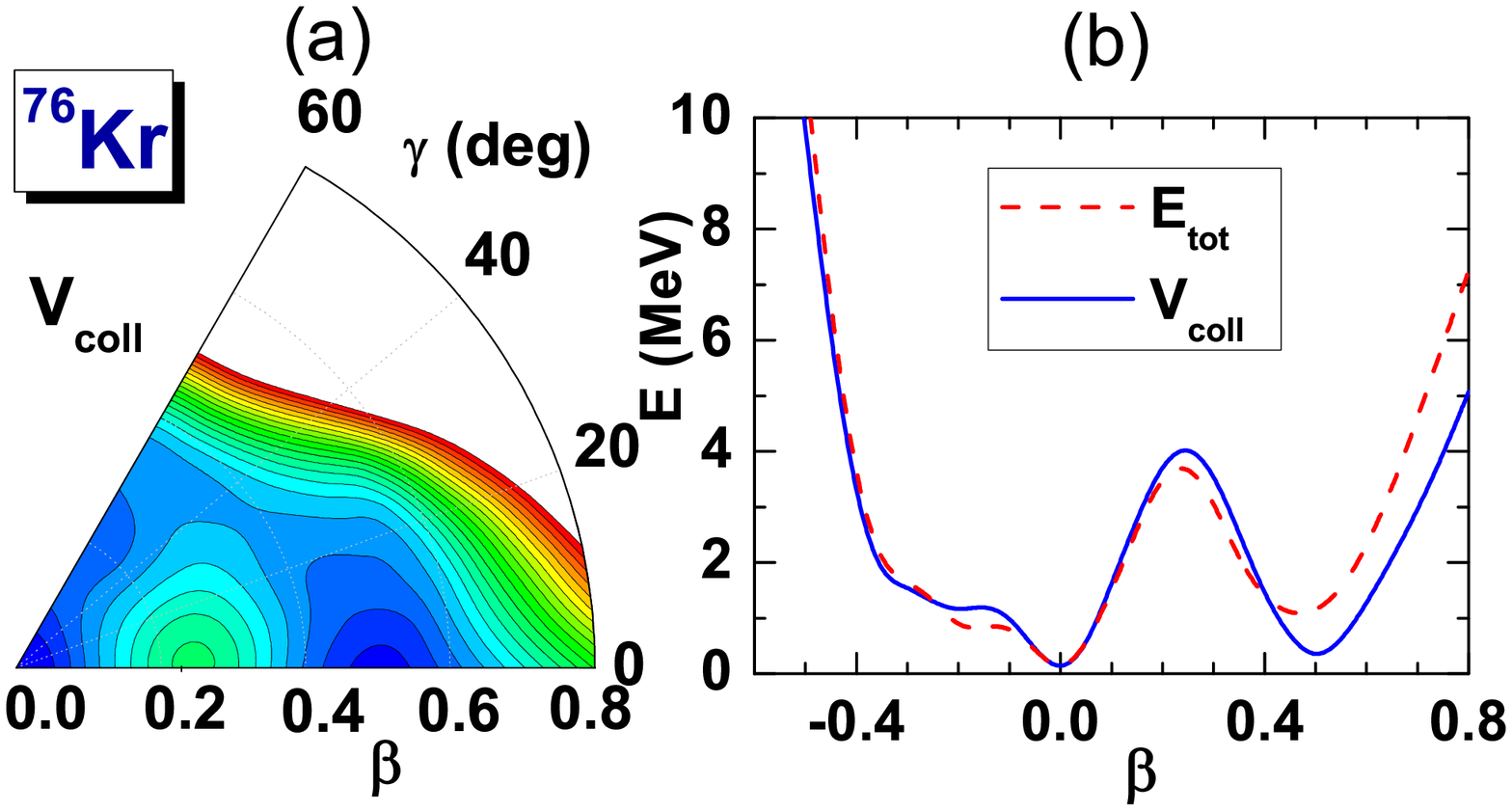}
\caption{(Color online) (a) Collective potential energy surface of $^{76}$Kr in the $\beta$-$\gamma$ plane [cf. Eq.(\ref{Vcoll})]. (b) Comparison of total energy surface and collective potential energy surface for $^{76}$Kr. All energies are normalized to the spherical shape.}
\label{Kr76_PES}
\end{figure}

 In this subsection, we will discuss the effects of dynamic correlations and triaxiality in detail by taking $^{76}$Kr as an example. Figure~\ref{Kr76_PES} displays the collective potential energy surface, subtracted the ZPEs of vibrational and rotational motions from the total energy, cf. Eq.(\ref{Vcoll}). It shows that the ZPEs deepen the prolate deformed minimum and reduce its energy difference with the spherical shape by about 0.9 MeV, which leads to a coexistence picture of competing spherical and prolate minima in the energy surface. Since the prolate minimum is broader, the ground-state of $^{76}$Kr is finally dominated by the large prolate deformed configurations in the 5DCH calculations.

\begin{figure}[]
\centering
\includegraphics[width=7.5cm]{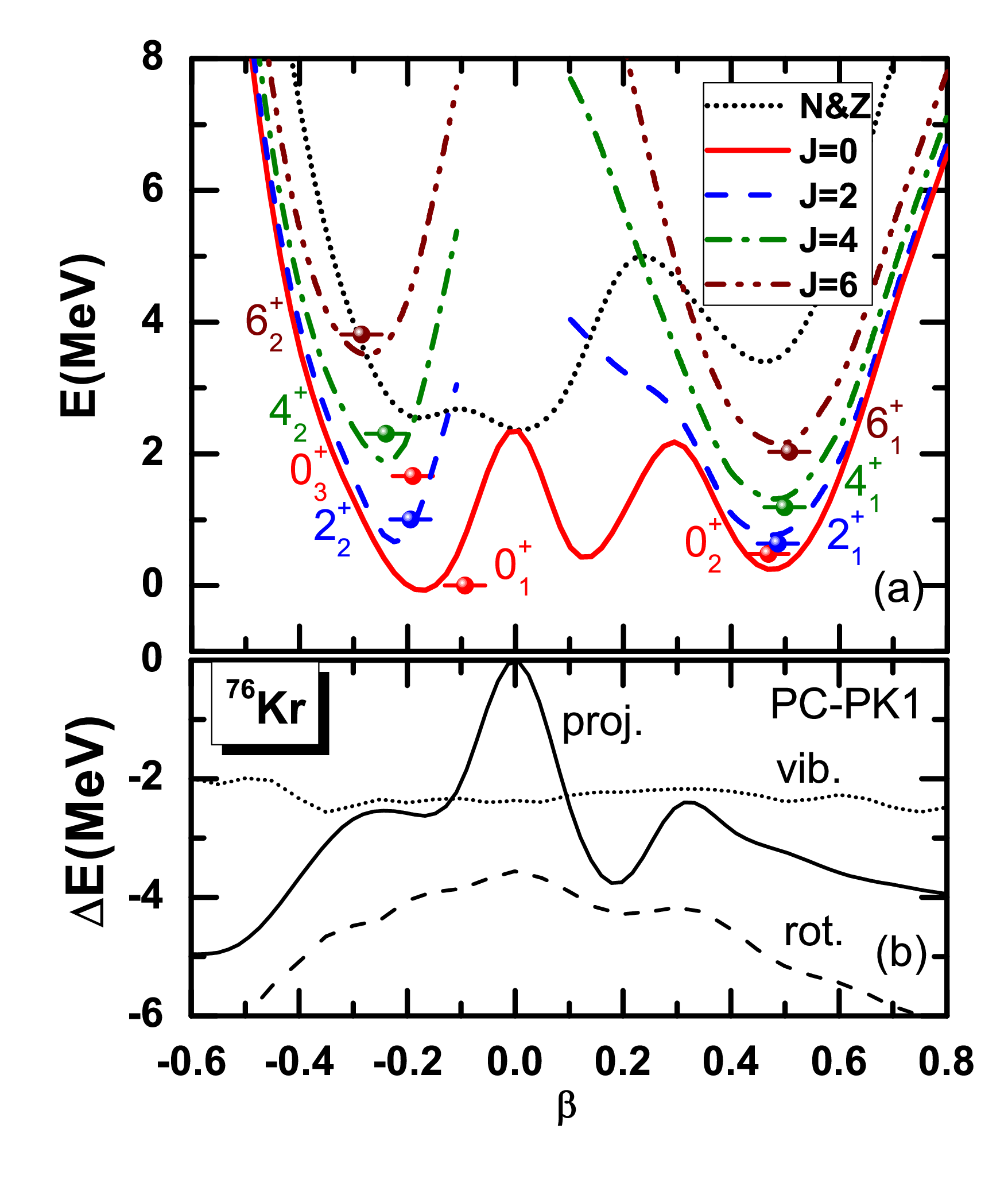}
\caption{(Color online) (a) Projected energy curves and low-lying states from the GCM+PNP+1DAMP calculations for  $^{76}$Kr. The bullets correspond to the lowest GCM solutions, which are placed at their average deformation. cf.~\cite{Yao10}. (b) The energy gained from projection on $J=0$: $\Delta E (\beta)= E_{\rm J=0}(\beta)- E_{\rm NZ}(\beta)$ (proj.), in comparison with the ZPEs of rotation motion $-\Delta V_{\textnormal{rot}}$ (rot.) and vibration motion $-\Delta V_{\textnormal{vib}}$ (vib.) evaluated with the cranking approximation. All the results are calculated using the relativistic mean-field wave functions by the PC-PK1 plus zero-range pairing force.}
\label{Rel:pes}
\end{figure}

\begin{figure}[]
\centering
\includegraphics[width=9cm]{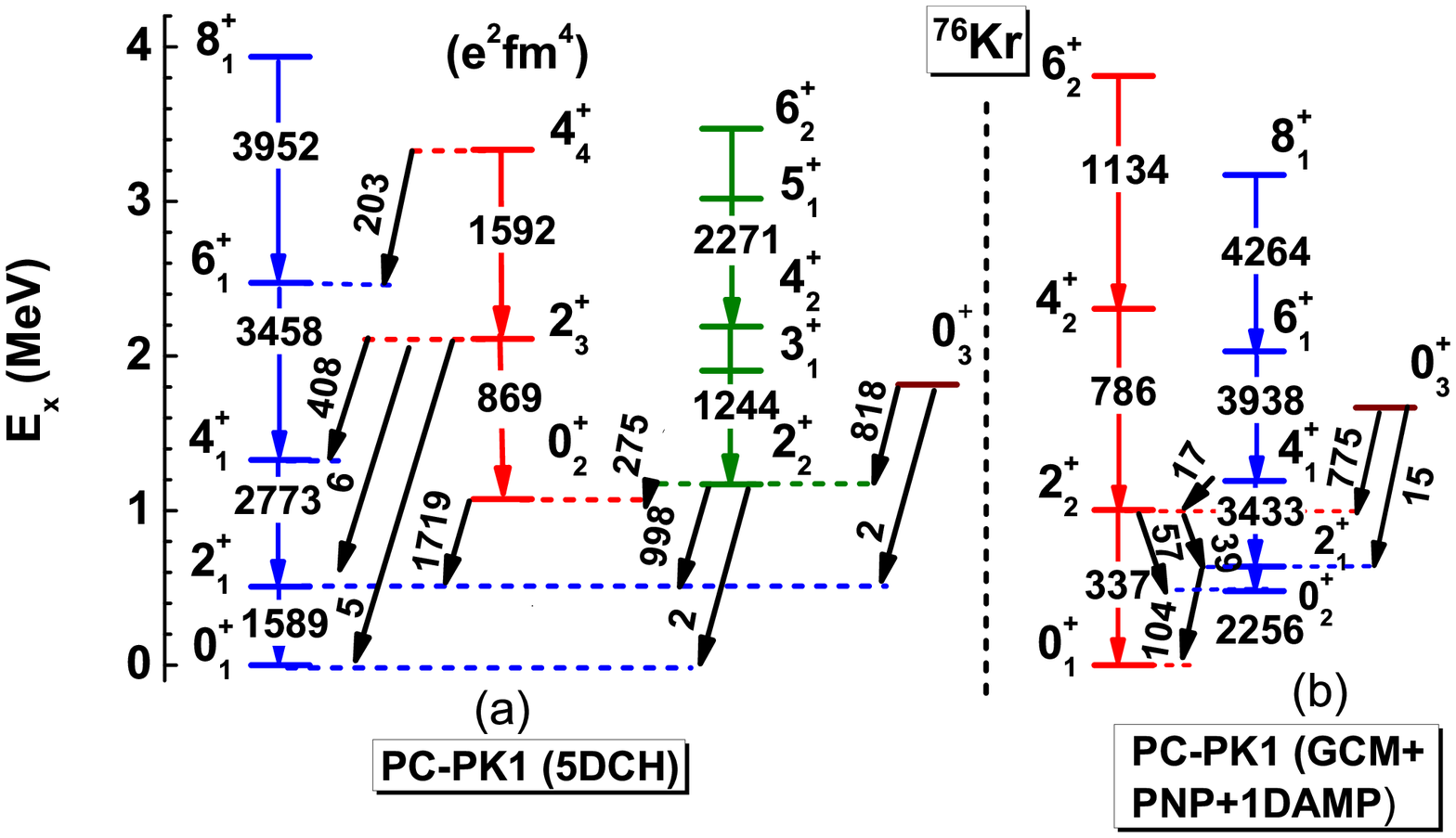}
\caption{(Color online) Comparison of low-spin spectra from (a) the 5DCH calculations and (b) the GCM+PNP+1DAMP calculations using the PC-PK1 plus density-independent zero-range pairing force for $^{76}$Kr.}
\label{Rel:com1}
\end{figure}
\begin{figure}[]
\centering
\includegraphics[width=9cm]{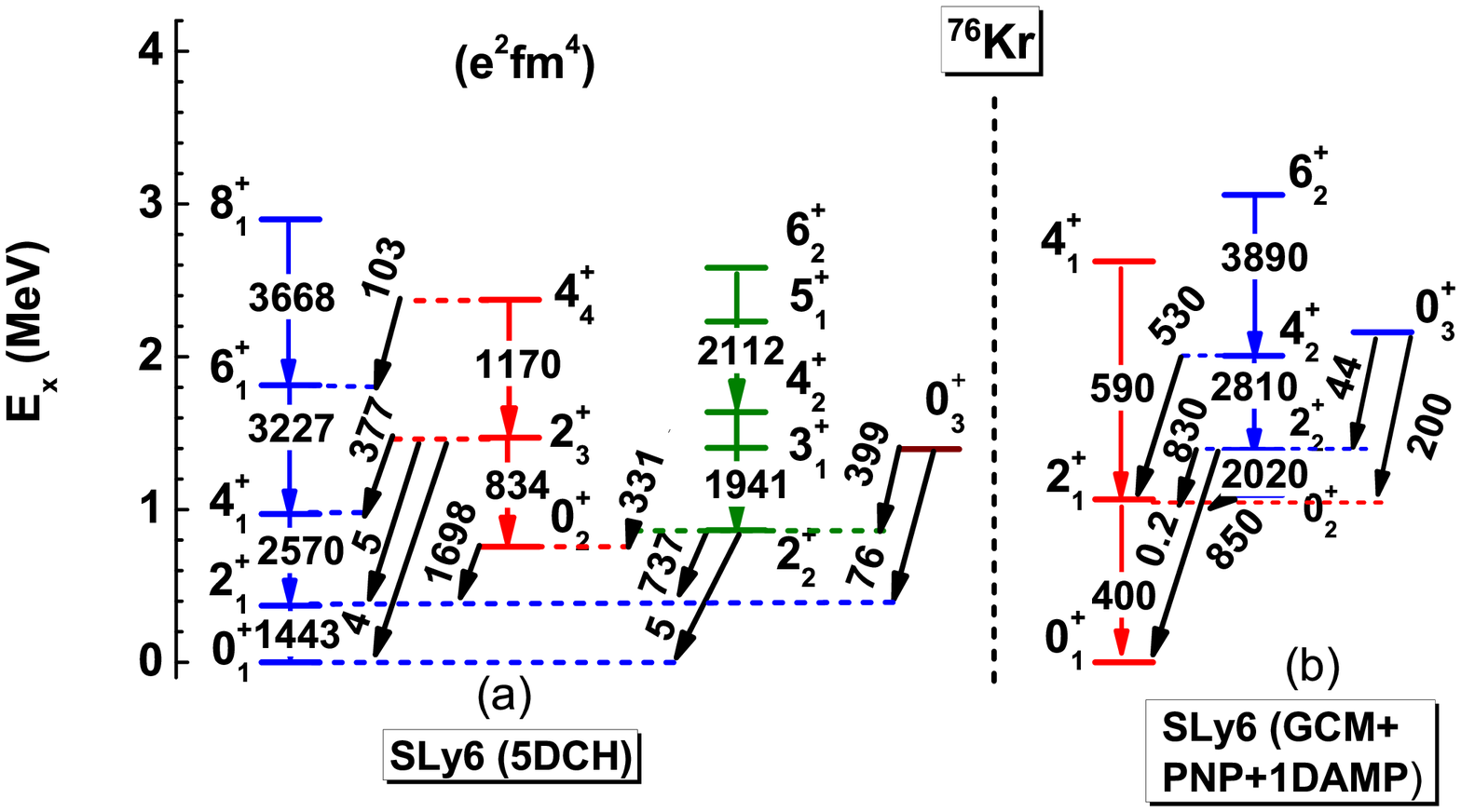}
\caption{(Color online) Same as Fig.~\ref{Rel:com1}, but using the SLy6 force plus density-dependent zero-range pairing force. The results of the GCM+PNP+1DAMP calculations are taken from Ref.~\cite{Bender06}.}
\label{Rel:com2}
\end{figure}

 In contrast to the 5DCH results, a weak $E2$ transition strength $B(E2:2^+_1\to 0^+_1)$ was obtained in the GCM+PNP+1DAMP calculation of Ref.~\cite{Bender06} for $^{76}$Kr using the SLy6 force due to the weakly oblate character of predicted ground-state band.

  Figure~\ref{Rel:pes} displays the energy curves projected on particle number and different angular momentum, calculated using the PC-PK1 force for the effective Lagrangian and a density-independent zero-range $\delta$ force for the pairing channel~\cite{Zhao10}. This calculation is carried out using our recently developed code~\cite{Yao10}, implemented with particle number projection. The pfaffian method is adopted to determine the norm overlaps~\cite{Robledo09}. The numerical details of GCM combined with PNP+1(3)DAMP techniques based on the relativistic mean-field wave functions will be given in a forthcoming paper~\cite{Yao13}. Similar results as those by the non-relativistic calculations in Ref.~\cite{Bender06} are obtained. The spherical configuration is purely a \mbox{$J=0$} configuration; hence, the projection for \mbox{$J=0$} does not gain any energy. A small deformation of the mean-field is sufficient to introduce higher $J$ components. The projection for \mbox{$J=0$} gives then an energy gain, which increases rapidly to reach about $2.7 (3.7)$ MeV around $\beta\approx -0.15 (0.20)$, as shown in Fig.~\ref{Rel:pes}. As a result, the weakly oblate deformed global minimum, together with a new weakly prolate deformed minimum, shows up in the $J=0$ projected energy surface.  At larger deformation, the energy gain still increases but at a slower rate, which brings down the large prolate deformed minimum around $\beta=0.5$. Similar features in the rotational correction energy have been observed in the neutron-deficient nuclei in lead region~\cite{Duguet03,Bender04,Guzman04,Yao12}.
 However, the energy gain from the projection for $J=0$ is not enough to make the large prolate deformed configuration lower in energy than the weakly oblate configuration. Therefore, a weakly deformed ground-state, instead of a large prolate deformed one, is also obtained in the relativistic GCM+PNP+1DAMP calculation.

 The main difference between our relativistic GCM+PNP+1DAMP calculation and the non-relativistic one is that the large deformed prolate configuration ($\beta\approx0.5$) here is much lower in energy and becomes yarst state at $J=2$, in comparison with $J=4$ for the non-relativistic one. In both calculations, the ground-state is dominated by the weakly oblate state, contradicting with the indication of data. As a result, the $B(E2:2^+_1\to 0^+_1)$ values in both calculations are much smaller than the data.

  Figures~\ref{Rel:com1} and \ref{Rel:com2} display the comparison of low-spin spectra from the 5DCH calculations and GCM+PNP+1DAMP calculations using PC-PK1 (with $\delta$ pairing force) and the SLy6 force respectively. Similar results are found for different forces in either 5DCH or GCM+PNP+1DAMP calculations . In the 5DCH calculations, the energy order of the oblate and prolate states is correctly reproduced and the calculated electric transition strengths are in good agreement with the data. These results are consistent with those in the 5DCH calculations using the Gogny D1S force~\cite{Girod09} and confirm the important role of triaxiality.

  We note that the dynamic correlation energies by the projections and GCM are different from those given by the cranking approximation. In Fig.~\ref{Rel:pes}, the ZPEs of rotational and vibrational motions by the cranking approximation are plotted for comparison. It is shown that the ZPE of vibrational motion is almost constant against the deformation parameter $\beta$, while the ZPE of rotational motion increases smoothly with the $\beta$. Compared with the energy gained from the exact AMP, the ZPE of rotational motion does not equal zero for the spherical state and the shell effect is not evident. As a result, the collective potential $V_{\rm coll}$, subtracted the ZPEs of rotational and vibrational motions from the total energy, does not present the weakly deformed oblate and prolate minima with $\vert\beta\vert$ in between 0.1 and 0.2. From this point of view, the 5DCH calculation is easier to give a large prolate deformed ground-state than the exact projected GCM calculation.

 We note that in the previous studies~\cite{Reinhard87,Hagino02}, the calculation of rotational correction energy using the Gaussian overlap approximation or a so called topologically invariant Gaussian overlap approximation has already been performed. In particular, the rotational correction energy from an exact AMP calculation was obtained in Ref.\cite{Guzman00}, which demonstrated that the exact restoration of the rotational symmetry is fundamental for a qualitative and quantitative description of the rotational energy. Therefore, an exact GCM calculation combined with the PNP+3DAMP techniques is required to pin down the triaxiality effect. The previous 3DAMP studies for light nuclei~\cite{Bender08,Yao10,Yao11-C} indicate that the AMP can lower the energy of the triaxial states. We expect that a triaxial global minimum with $\vert\beta\vert$ in between 0.2 and 0.5 may show up in the triaxial projected energy surface with $J=0$, and modify the conclusion drawn from the GCM calculations restricted to the axial case.

%
\section{summary}\label{Sec.IV}
%

In summary, we have presented a systematical beyond RMF study of the low-lying states in the neutron-deficient krypton isotopes. The excitation energies and electric multipole transition strengths have been obtained by solving a five-dimensional collective Hamiltonian (5DCH) with parameters determined from the RMF calculations using the PC-PK1 force. The results have been compared with those obtained by the similar calculations but using the Gogny D1S and the Skyrme SLy6 forces as well as by the exact GCM+PNP+1DAMP calculations. The results of the 5DCH calculations based on different types of forces are similar, except difference in the size of configuration mixing. We find a picture of oblate-triaxial-prolate shape transition in the neutron-deficient Kr isotopes. Coexistence of low-lying excited $0^+$ states has been shown to be a common feature in the nuclei of this mass region.

The results from the beyond mean-field calculations using the relativistic PC-PK1 force, together with the calculated results using the SLy6 force, confirm the conclusion given in Ref.~\cite{Girod09} that triaxiality is important in reproducing the energy order of the oblate and prolate states. Moreover, we have analyzed the role of the intruder $1g_{9/2}$ orbital in preventing a strong mixing between the large prolate deformed configurations with the weakly oblate deformed ones, which serves as a mechanism responsible for the shape coexistence phenomenon in the Kr isotopes around $N=40$. Furthermore, we have illustrated the important role of dynamic correlations on the onset of large collectivity in $^{76}$Kr.

The CDFT provides a self-consistent and universal description of nuclei all over the nuclear chart. This study provides an example that the CDFT can describe not only the nuclei with a single-configuration dominated structure, but also the nuclei with a coexistence structure of distinctly different shapes in low-excitation energy and the transition behavior from one dominate quadrupole shape to another when the effects of dynamic correlations and triaxiality are considered properly. Although the model of 5DCH is able to describe the systematics of excitation energies and transition strengths in the neutron-deficient Kr isotopes, the exact GCM calculations combined with the PNP+3DAMP techniques are highly demanded to pin down the importance of triaxiality. As the dynamic correlations from the ZPEs are different in detail from those by the exact projected GCM calculations, the conclusions drawn from the 5DCH studies might be altered. Work along this direction is in progress.

\begin{acknowledgements}

This work was supported in part by the Major State 973 Program 2013CB834400,
the NSFC under Grant Nos. 10975008, 10947013, 11175002, 11105110, and 11105111, the Research
Fund for the Doctoral Program of Higher Education under Grant
No. 20110001110087, the European Union's Seventh Framework Programme ENSAR under grant agreement n262010, the Fundamental Research Funds for the Central Universities
(XDJK2010B007 and DJK2011B002) and the NSF of Chongqing cstc2011jjA0376.
\end{acknowledgements}

\end{document}